\documentclass[preprint,aps,prl,showpacs,preprintnumbers, superscriptaddress, amsmath,amssymb]{revtex4}

\usepackage{stmaryrd} 
\usepackage{mathtools} 
\usepackage{stmaryrd} 

\usepackage{amsmath}
\usepackage{amsfonts}
\usepackage{amssymb}
\usepackage{graphicx}%

\begin{document}

\author{Renan Cabrera}
\email{rcabrera@princeton.edu}
\affiliation{Department of Chemistry, Princeton University, Princeton, NJ 08544, USA} 

\author{Denys I. Bondar}
\affiliation{Department of Chemistry, Princeton University, Princeton, NJ 08544, USA} 

\author{Herschel A. Rabitz}
\affiliation{Department of Chemistry, Princeton University, Princeton, NJ 08544, USA} 

\title{NMR quantum gate factorization through canonical cosets  }

\date{\today}

\begin{abstract}
The block canonical coset decomposition is developed as a universal algorithmic tool to synthesize  n-qubit quantum gates out of experimentally realizable NMR elements. The two-, three-, and four-qubit quantum Fourier transformations are worked out as examples. The proposed decomposition bridges the state of the art numerical analysis with NMR quantum gate synthesis.  
\end{abstract}


\maketitle

\emph{Introduction.}
According to the quantum computing circuit model, a quantum gate is a unitary operator designed to perform a predefined operation. The primary aim is to synthesize unitary operators suitable for experimental implementation.
Currently, quantum gate synthesis is achieved either by factorization methods, 
as in the strategy ``divide and conquer,'' or by optimal control theory \cite{Schulte-Herbruggen2005a,SCHULTE-HERBRUGGEN2010,Moore2011}.

The factorization methods' development was initiated by utilizing the KAK Cartan decomposition to attain minimal time two-qubit operations \cite{khaneja2001time, khaneja2001cartan, Dirr2006} with subsequent extensions to multi-qubit gates \cite{khaneja2001cartan,DAlessandro2007}.
However, the Cartan decomposition, coming from Lie group theory, was not constructive and actual applications required 
the independent development of numerical routines \cite{earp2005constructive, dagli2008general}. Inspired by numerical analysis, alternative approaches, built upon the cosine-sine matrix factorization \cite{mottonen2004quantum,bullock2004canonical,vartiainen2004efficient,Nakajima2006,Shende2006,DAlessandro2007},
were employed to constructively decompose an arbitrary quantum operation in terms of single-qubit and CNOT gates as well as to estimate the complexity of such implementations.
 Since CNOT gates are not most convenient in NMR setups \cite{ZhigangZhangGoongChen2009}, these developments did not find experimental applications. While NMR two-qubit synthesis has been thoroughly explored \cite{khaneja2001time,Zhang2004,Zhang2005}, 
multi-qubit factorization remains largely an open  problem \cite{Khaneja2007,Yuan2011,carlini2011time}.

In this Letter, we introduce  the canonical coset decomposition as a constructive  factorization technique to build quantum gates out of standard single- and two-qubit NMR operations. To illustrate this method's ability, we present the explicit factorizations 
of the two-, three-, and four-qubit quantum Fourier transform, which is the basic component of many quantum computing algorithms.

\emph{Canonical Coset Decomposition} is a general group-theoretic tool originally devised to  study the geometric structure of unitary matrices \cite{gilmore1973lgl,gilmore2008lgp}. 
In particular, this method allowed to obtain analytic expressions for the measure of unitary operators
\cite{gilmore1973lgl,gilmore2008lgp} as well as the measure \cite{Cabrera2009} and metric \cite{Akhtarshenas2007} of mixed quantum states.

The canonical coset decomposition of a unitary matrix $U$ is
\begin{align}
U &=  \left(
    \begin{array}{c|c} 
  \sqrt{\mathbf{1}  -  X^\dagger X} &  -X^\dagger \\ \hline
   X & \sqrt{ \mathbf{1} - X X^\dagger  }
   \end{array}\right)
    \left( 
   \begin{array}{c|c} V_1 & 0 \\ \hline
   0 & V_2
   \end{array}\right),
\label{CCC}
\end{align}
with
\begin{align}
   U   &=
  \left(    \begin{array}{c|c} 
  U_{11} &  U_{12} \\ \hline
   U_{21} & U_{22}   \end{array}\right), \\
  X &= U_{21} [\mathbf{1}  -  U_{21}^\dagger U_{21}]^{1/2} U_{11}^{-1},\\
  V_1 &=  [\mathbf{1}  -  X^\dagger X]^{1/2} U_{11} + X^\dagger U_{21}, \\
  V_2 &=  [\mathbf{1}  -  X X^\dagger ]^{1/2} U_{22} - X U_{12},
\label{XA1A2}
\end{align}
where $U_{11}$ and $ [\mathbf{1}  -  U_{21}^\dagger U_{21}]^{1/2} $ are invertible. Throughout the Letter we exclusively employ the \emph{block canonical coset decomposition}, which is a special case of Eq. (\ref{CCC}) with $U_{ij}$, $X$, and $V_{1,2}$ being square matrices of the same dimension. 

The block canonical coset decomposition is a a very powerful tool because each term in the r.h.s. of Eq.  (\ref{CCC}) is simpler than the original matrix $U$. In particular, the first term can be expressed for  $X =  \frac{\sin \sqrt{B B^\dagger}}{\sqrt{BB^\dagger} } B$ \cite{gilmore1973lgl} as
\begin{align}
      \left(    \begin{array}{c|c} 
  \sqrt{\mathbf{1}  -  X^\dagger X} &  -X^\dagger \\ \hline
   X & \sqrt{ \mathbf{1} - X X^\dagger  }
   \end{array}\right) = 
  \exp{ 
 \left( 
   \begin{array}{c|c} 0 & -B^\dagger \\ \hline
   B & 0
   \end{array}\right)
  };
\label{exp-p}
\end{align}
additionally, the second term forms a subgroup. In other words, the Lie algebra $\mathcal{L}(U)$ of unitary matrices is decomposed into 
a block diagonal subalgebra $\mathfrak{t}$ and its complement $\mathfrak{p}$
\begin{align}
  \mathcal{L}(U) = \mathfrak{p} \oplus \mathfrak{t} \equiv
   \left( 
   \begin{array}{c|c} 0 & -B^\dagger \\ \hline
   B & 0
   \end{array}\right)
 \oplus
  \left( 
   \begin{array}{c|c} \log V_1 & 0 \\ \hline
   0 & \log V_2
   \end{array}\right).
\end{align}
For n-qubit gates $U(2^n)$, the factorization (\ref{CCC}) reads
\begin{align}
\label{ccc-decomposition}
 U(2^n) = \frac{U(2^n)}{U(2^{n-1}) \otimes  U(2^{n-1})} [ U(2^{n-1}) \otimes U(2^{n-1})],  
\end{align}
where the right factor corresponds to the block diagonal subgroup.

The canonical coset decomposition is closely related to other techniques. In particular, if $X$ is diagonal then the first term in the r.h.s. of Eq.  (\ref{CCC}) is equal to the middle term of the cosine-sine decomposition \cite{mottonen2004quantum}. When $X$ is a single column, the decomposition (\ref{CCC}) is equivalent to the Householder transformation \cite{cabrera2010canonical}, which is not only one of the most 
important operations in numerical analysis \cite{cipra2000best}, but also very useful for the generation of n-level unitary operators \cite{Bullock2005,ivanov:022323,ivanov:012335}.

\emph{Two-qubit gates}. 
According to Eq. (\ref{ccc-decomposition}), the $U(4)$ group associated with two-qubit 
gates is decomposed as
\begin{align}
\label{U4-2}
  U(4) = \frac{U(4)}{U(2)\otimes U(2)  } [ U(2)\otimes U(2) ].
\end{align}
The reverse form of Eq. (\ref{U4-2}),
\begin{equation}
\label{U4-2-left}
  U(4) =  U(2)\otimes U(2)  \frac{U(4)}{U(2)\otimes U(2)  }, 
\end{equation}
can be achieved by using the identity $ U = (U^{\dagger})^{\dagger} $
and performing the decomposition (\ref{U4-2}) on $U^\dagger$.

The matrix $\log$ of $U(4)$ is spanned by the basis elements listed in 
Tables \ref{Table1}(A) and \ref{Table1}(B), where the basis of 
the subalgebra $\mathfrak{t}$ is given in Table \ref{Table1}(B). 
\begin{table}
\begin{center}
  \begin{tabular}{c | c| c| c| c| } 
     \hline 
    (A) & $\sigma_1\otimes\sigma_0$ &  $\sigma_1\otimes\sigma_1$ &   $\sigma_1\otimes\sigma_2$  &   $\sigma_1\otimes\sigma_3$ \\ 
    & $\sigma_2\otimes\sigma_0$ &  $\sigma_2\otimes\sigma_1$ &  $\sigma_2\otimes\sigma_2$  &  $\sigma_2\otimes\sigma_3$ \\ \hline
     \hline
    (B) & $\sigma_0\otimes\sigma_0$  &  $\sigma_0\otimes\sigma_1$  & $\sigma_0\otimes\sigma_2$   &   $\sigma_0\otimes\sigma_3$   \\ 
    & $\sigma_3\otimes\sigma_0$   &  $\sigma_3\otimes\sigma_1$   & $\sigma_3\otimes\sigma_2$   &  $\sigma_3\otimes\sigma_3$  \\ 
    \hline
  \end{tabular}
\end{center}
 \caption{ Basis elements for (A) the complement $\mathfrak{p}$  and (B) the subalgebra $\mathfrak{t}$ of the $u(4)$ Lie algebra.}\label{Table1}
\end{table} 
Since $\mathfrak{t}$ includes both single and two-qubit interactions, isolation of single-qubit 
operations necessities the following factorization:
\begin{align}
 \label{U2-isolation}
  U(2)\otimes U(2) &= \frac{ U(2)\otimes U(2)  }{  U(2)  }   U(2), \\
  U(2)\otimes U(2) &= 
  \left( 
   \begin{array}{c|c}  G_1 & 0 \\ \hline
   0 &  G_2
   \end{array}\right), \label{subgroup-U2} \\
 U(2) &=  \left( 
   \begin{array}{c|c}  S_1 & 0 \\ \hline
   0 &  S_1
   \end{array}\right), \label{single-U2} \\
  \frac{ U(2)\otimes U(2)  }{  U(2)  } &= 
 \left( 
   \begin{array}{c|c}  S_2 & 0 \\ \hline
   0 &  S_2^{\dagger}
   \end{array}\right), \label{coset-U2}
\end{align}
where  $S_2 = [G_1G_2^\dagger]^{1/2}$ and $S_1 = S_2^{\dagger} G_1 $.  Equation (\ref{single-U2}) accomplishes
 the desired isolation. Even thought Eq. (\ref{coset-U2}) is not a single-qubit operator, it can be readily 
transformed into one by employing the standard NMR methods. Thus, 
making the factorization (\ref{U2-isolation}) suitable for NMR experiments.

The first factor in the r.h.s. of Eq. (\ref{U4-2}) can be further simplified by a recursive 
factorization (\ref{U4-2}) and (\ref{U4-2-left})
\begin{align}
 \label{recursive-decomposition-U4}
 U(4) =  U(2)\otimes U(2)  \frac{U(4)}{ [U(2)\otimes U(2)]^2  } U(2) \otimes U(2),
\end{align}
where the middle term becomes simple enough for NMR synthesis.
Having initiated the 
unitary operators
\begin{align}
\label{pivot-pi-half}
V \coloneqq \exp\left( i \frac{\pi}{4} \sigma_2 \otimes \sigma_0  \right), \quad
\widetilde{U^{(2)}} \coloneqq \mathbf{1}, \quad \widetilde{W^{(2)}} \coloneqq \mathbf{1},
\end{align}
the recursive decomposition is realized by the following 
algorithm:
\begin{enumerate}
 \item Perform the factorization (\ref{U4-2}) of $U$  and
  assign $U^{(1)} \coloneqq  \frac{U(4)}{U(2)\otimes U(2)  }$, 
    $U^{(2)} \coloneqq U(2) \otimes U(2) $;
 \item  $\widetilde{U^{(2)}} \coloneqq U^{(2)} \widetilde{U^{(2)}}$, ${U^{\prime}}^{(1)} \coloneqq V^\dagger U^{(1)} V  $;
 \item Perform the decomposition  (\ref{U4-2-left}) of 
${U^{\prime}}^{(1)}$
 and assign $W^{(1)} \coloneqq V [ U(2)\otimes U(2) ] V^{\dagger} $, 
 $W^{(2)} \coloneqq  V \frac{U(4)}{U(2)\otimes U(2)} V^\dagger $;
 \item $\widetilde{W^{(1)}} \coloneqq \widetilde{W^{(1)}}W^{(1)}$, $ U \coloneqq W^{(2)} $;
 \item Return to step 1 if the convergence conditions $U^{(2)} \approx \mathbf{1}$ and $W^{(1)}  \approx \mathbf{1}$ are not met; 
 \item Finally, the right subgroup in Eq. (\ref{recursive-decomposition-U4})
is stored in  $\widetilde{U^{(2)}}$ and the left subgroup --  $\widetilde{W^{(1)}}$. The  middle term $U(4) / [U(2)\otimes U(2)]^2$ is in $W^{(2)}$.
\end{enumerate}
The matrix $\log$ of the left, middle, and right terms in Eq. (\ref{recursive-decomposition-U4}) are spanned by the elements 
from Tables \ref{Table2}(A), \ref{Table2}(B), and \ref{Table2}(C), respectively.  
Since the middle term is made of the elements $\{ \sigma_2 \otimes \sigma_i \}_{i=0,1,2,3}$, it is well suited for NMR synthesis. 

Importantly, the developed algorithm is applicable to an arbitrary n-qubit system after adjusting the definition of $V$ according to
\begin{align}\label{pivotV}
	V \coloneqq \exp\Big( i \frac{\pi}{4} \sigma_2 \otimes  
		\underbrace{\sigma_0 \otimes \cdots \otimes \sigma_0}_{\mbox{$(n-1)$ times}}  \Big).
\end{align}   
 
\begin{table}
\begin{center}
  \begin{tabular}{c | c| c| c| c| }
     \hline
       (A) &  $\sigma_0\otimes\sigma_1$    &   $\sigma_0\otimes\sigma_2$   &   $\sigma_0\otimes\sigma_3$ &   \\ 
     & $\sigma_2\otimes\sigma_0$  &  $\sigma_2\otimes\sigma_1$    &
      $\sigma_2\otimes\sigma_2$   &   $\sigma_2\otimes\sigma_3$   \\ \hline \hline
     (B) & $\sigma_2\otimes\sigma_0$ &  $\sigma_2\otimes\sigma_1$ &   $\sigma_2\otimes\sigma_2$  &   $\sigma_2\otimes\sigma_3$ \\ \hline\hline
      (C) &$\sigma_0\otimes\sigma_0$ &  $\sigma_0\otimes\sigma_1$ &  $\sigma_0\otimes\sigma_2$  &  $\sigma_0\otimes\sigma_3$ \\  
      & $\sigma_3\otimes\sigma_0$   &  $\sigma_3\otimes\sigma_1$   & $\sigma_3\otimes\sigma_2$   &  $\sigma_3\otimes\sigma_3$   \\
    \hline
  \end{tabular}
\end{center}
 \caption{ Basis elements for  the Lie algebra $u(4)$ corresponding to (A) left, (B) middle, (C) right terms in Eq. (\ref{recursive-decomposition-U4}).}\label{Table2}
\end{table}

\begin{figure}[ht] 
\centering
\includegraphics[scale=0.8]{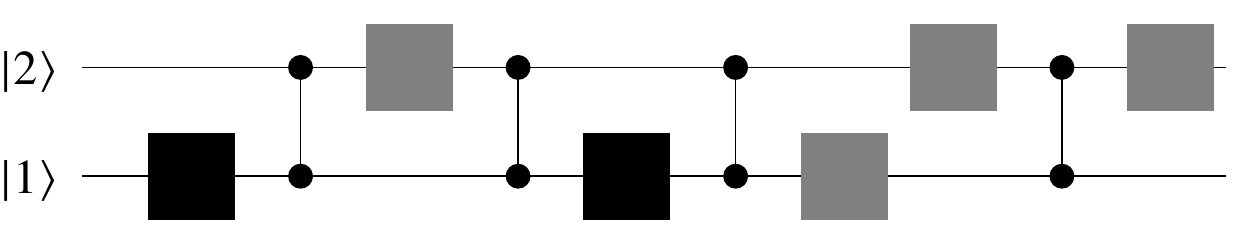}
\caption{The schematic representation of the two-qubit Fourier matrix factorization (\ref{F4-sequence}).
Black boxes represent single-qubit $\pi/4$ rotations; 
gray boxes -- taylored single-qubit operators specified by Eq. (\ref{single-qubits-F4}).} 
\label{figure-1}
\end{figure}

\emph{The two-qubit quantum Fourier transform} $F$
is used to illustrate the proposed algorithm. As shown in Sec. I of Ref. \cite{SupplementalM1}, 
the recursive decomposition (\ref{recursive-decomposition-U4}) leads to
\begin{align}
 F =& \exp\left(-\frac{i\pi}{4} \sigma_1\otimes\sigma_0\right) \exp\left(-\frac{i\pi}{4} \sigma_3\otimes\sigma_3\right)  \nonumber \\
\times&  F_1\exp \left(\frac{i\pi}{4} \sigma_3\otimes\sigma_3 \right) \exp\left(\frac{i\pi}{4} \sigma_1\otimes\sigma_0 \right) \nonumber\\
\times&  \exp\left(-\frac{i\pi}{4} \sigma_3\otimes\sigma_3\right)F_2 F_{31} \exp\left(\frac{i\pi}{4} \sigma_3\otimes\sigma_3\right)U_{32},
\label{F4-sequence}
\end{align}
with 
\begin{align}
i \log(F_1)  = & \, 0.55536 \, \sigma_0 \otimes (\sigma_1 +\sigma_2), \nonumber\\
i \log(F_2)  = & \, -0.392699\, \sigma_2 \otimes \sigma_0, \nonumber\\
i \log(F_{31}) = & \, 0.55536\, \sigma_0 \otimes (\sigma_1 - \sigma_2 ) ,\nonumber   \\
i \log(U_{32})= & \, \sigma_0 \otimes (-1.1781\sigma_{0} + 0.785398 \sigma_3).   
\label{single-qubits-F4}
\end{align}
The pictorial representation of this factorization is shown in Fig. \ref{figure-1}.

\emph{Three-qubit gates.}
According to Eq. (\ref{ccc-decomposition}), a three-qubit gate is decomposed as
\begin{equation}
  U(8) =  \frac{U(8)}{U(4) \otimes U(4)} [U(4)\otimes U(4)] 
\label{U8-decomp-2},
\end{equation}
where $U(4) \otimes U(4)$ is block diagonal. 
The two-qubit decomposition (\ref{recursive-decomposition-U4}) is extended to three-qubit 
gates 
\begin{align}
 \label{recursive-decomposition-U8}
 U(8) =  U(4)\otimes U(4)  \frac{U(8)}{ [U(4)\otimes U(4)]^2  } U(4) \otimes U(4),
\end{align}
where the central factor's matrix $\log$ is spanned by $\{\sigma_{2ij} \}_{i,j=0,1,2,3}$, where $\sigma_{ijk} = \sigma_{i} \otimes \sigma_j \otimes \sigma_{k}$.
According to Eq. (\ref{pivotV}), the factorization algorithm is applicable in the three-qubit case once the definition of $V$ in Eq. (\ref{pivot-pi-half}) is replaced by  $V \coloneqq \exp( i\pi \sigma_{200} / 4 )$.
The decomposition (\ref{U2-isolation}) is naturally generalized 
to  isolate the subgroup $U(4)$ associated 
with two-qubits
\begin{align}
 \label{U4-isolation}
  U(4)\otimes U(4) = \frac{ U(4)\otimes U(4)  }{  U(4)  }   U(4). 
\end{align}
where the matrix logarithms of $U(4)$ and $U(4)\otimes U(4)  / U(4)$ are linear combinations of $\{ \sigma_{0ij} \}_{i,j=0,1,2,3}$ and  $\{ \sigma_{3ij} \}_{i,j=0,1,2,3}$, respectively.

\begin{figure}[ht] 
\centering
\includegraphics[scale=0.8]{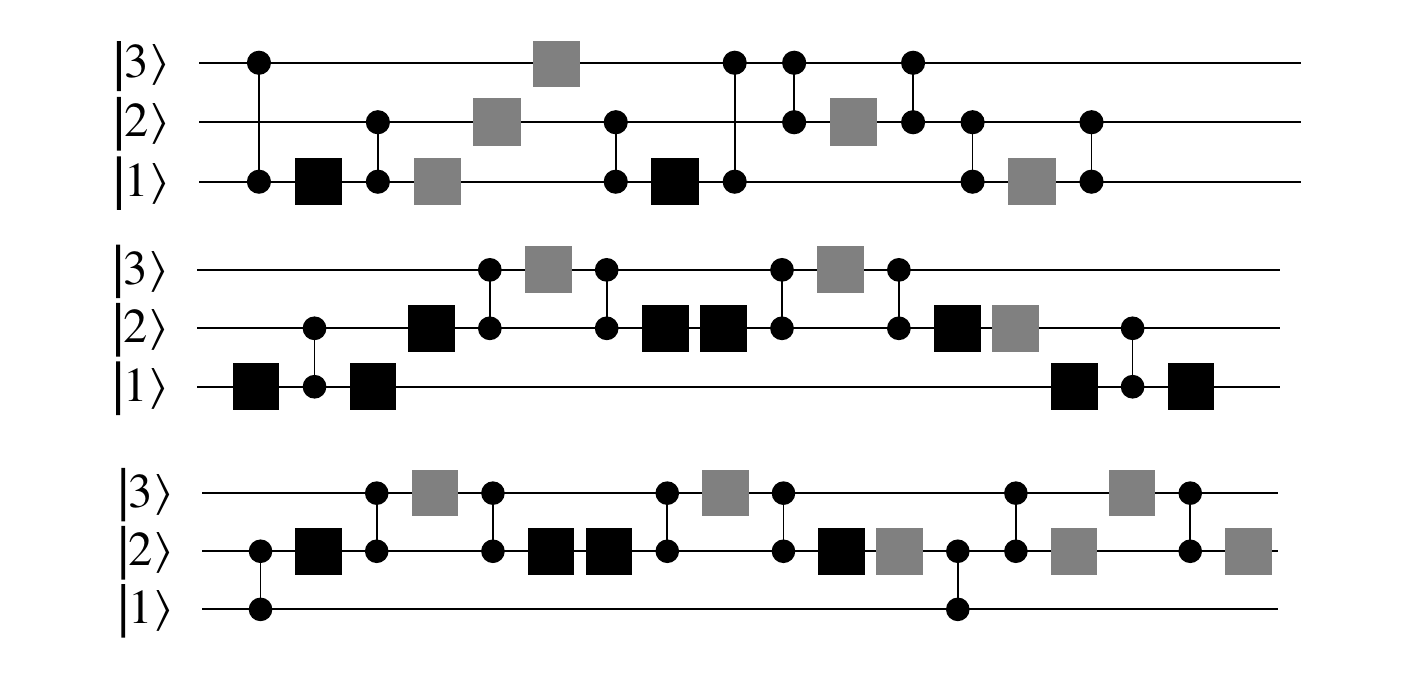}
\caption{ The schematic representation of the three-qubit Fourier matrix factorization from  Eqs. (\ref{F8-U1}), (\ref{F8-U2}), and (\ref{F8-U3}). 
Black boxes represent single-qubit $\pi/4$ rotations; 
gray boxes -- taylored single-qubit operators.
} 
\label{figure-2}
\end{figure}

The factorization (\ref{recursive-decomposition-U8})
of the \emph{three-qubit quantum Fourier transform}  results in $  F = U_1 U_2 U_3$.
Each of these factors are further decomposed in Sec. II of \cite{SupplementalM1}
\begin{align}
 U_1 =&  \exp \left(\frac{i\pi}{4} \sigma _{303}\right) 
 \exp \left(\frac{i\pi}{4} \sigma _{100}\right) \exp \left(\frac{i\pi}{4}  \sigma _{330}\right) F_{11}\nonumber\\ 
&\times \exp \left(-\frac{i\pi}{4} \sigma _{330}\right) \exp \left(-\frac{i\pi}{4} \sigma _{100}\right)
 \exp \left(-\frac{i\pi}{4}  \sigma _{303}\right) \nonumber\\
&\times\exp \left(\frac{i\pi}{4}  \sigma _{033}\right)F_{12}\exp \left(-\frac{i\pi}{4} \sigma _{033}\right)\nonumber \\ 
&\times \exp \left(\frac{i\pi}{4}  \sigma _{330}\right)F_{13} \exp \left(-\frac{i\pi}{4}  \sigma _{330}\right),
\label{F8-U1}\\
U_2 =&  \exp \left(\frac{i\pi}{4}  \sigma _{100}\right) \exp \left(\frac{i\pi}{4} \sigma _{330}\right)
 \exp \left( -\frac{i\pi}{4}  \sigma _{100}\right) \nonumber \\
 \times&\exp \left(\frac{i\pi}{4} \sigma _{020}\right)
 \exp \left(\frac{i\pi}{4} \sigma_{033} \right) F_{21}
\exp \left(- \frac{i\pi}{4} \sigma _{033}\right) \nonumber \\
\times& \exp \left(-\frac{i\pi}{4}  \sigma _{020}\right) 
\exp \left(\frac{i\pi}{4} \sigma _{010}\right) 
\exp \left(\frac{i\pi}{4}  \sigma _{033}\right) F_{22} \nonumber \\
\times& \exp \left(-\frac{i\pi}{4} \sigma_{033}\right) \exp \left(-\frac{i\pi}{4} \sigma _{010}\right)W_{23} 
\exp \left(\frac{i\pi}{4} \sigma _{100}\right) \nonumber \\
\times& \exp \left(-\frac{i\pi}{4} \sigma _{330}\right) 
\exp \left(-\frac{i\pi}{4}  \sigma _{100}\right),
\label{F8-U2}\\
 U_3  =& \exp \left(\frac{i\pi}{4} \sigma _{330}\right) 
        \exp \left(\frac{i\pi}{4} \sigma _{020}\right) 
\exp \left(\frac{i\pi}{4} \sigma _{033}\right) F_{311}\nonumber \\
   \times& \exp \left(-\frac{i\pi}{4}  \sigma_{033}\right) 
   \exp \left(-\frac{i\pi}{4}  \sigma _{020}\right)
   \exp \left(\frac{i\pi}{4}  \sigma _{010}\right) \nonumber \\
 \times& \exp \left(\frac{i\pi}{4}   \sigma _{033}\right)F_{312} 
  \exp \left(-\frac{i\pi}{4}  \sigma _{033}\right)
  \exp \left(-\frac{i\pi}{4}  \sigma _{010}\right) \nonumber \\ 
  \times& W_{313}  
  \exp \left(-\frac{i\pi}{4}  \sigma_{330}\right) 
  \exp \left(\frac{i\pi}{4}   \sigma _{033}\right) \nonumber \\
\times& F_{323}   \exp \left(-\frac{i\pi}{4} \sigma _{033}\right)U_{324}, 
\label{F8-U3}
\end{align}
where $F_{j\cdots}$, $W_{j\cdots}$ and $U_{jk\cdots}$ are single-qubit operators. This factorization is also depicted in Fig. \ref{figure-2}.

Finally, the four-qubit Fourier transform is elaborated in Sec. III of Ref. \cite{SupplementalM1}.

\emph{Conclusions}. 
The block canonical coset decomposition is introduced as an algorithmic procedure to synthesize an arbitrary n-qubit quantum gate out of one- and two-qubit NMR operations. In particular, we worked out the implementations of the two-, three-, and four-qubit quantum Fourier transform given by a dense matrix (i.e., without zero entries), therefore presenting a convincing illustration of our method's capability. Moreover, a highly optimized numerical implementation of the developed algorithm can be carried out through the block Householder decomposition \cite{Rotella1999,anderson1990lapack,bischof1987wy,urias2010}. Thus, the current work provides an important bridge between the state of the art numerical analysis and NMR quantum gate synthesis.

\end{document}